\documentstyle[psfig,12pt,aaspp4]{article}
 
\def\be{\begin{equation}}
\def\ee{\end{equation}}

\def\m{$\mu$m}
\def\HI{\ion{H}{1}}
\def\HII{\ion{H}{2}}
\def\NII{[\ion{N}{2}]}
\def\NIII{[\ion{N}{3}]}
\def\CII{[\ion{C}{2}]}
\def\OI{[\ion{O}{1}]}
\def\OIII{[\ion{O}{3}]}
\def\ISO{{\it ISO}}
\def\IRAS{{\it IRAS}}
\def\SIRTF{{\it SIRTF}}
\def\ISOcolor{${f_\nu (6.75 \mu {\rm m})} \over {f_\nu (15 \mu {\rm m})}$}
\def\IRAScolor{${f_\nu (60 \mu {\rm m})} \over {f_\nu (100 \mu {\rm m})}$}
\def\IRAScolorb{${f_\nu (12 \mu {\rm m})} \over {f_\nu (25 \mu {\rm m})}$}
\def\IRAScolorc{${f_\nu (25 \mu {\rm m})} \over {f_\nu (60 \mu {\rm m})}$}

\begin {document}
\title {The Infrared Spectral Energy Distribution of Normal Star-Forming Galaxies}
 
\author {Daniel A. Dale, George Helou, Alessandra Contursi, Nancy A. Silbermann, and Sonali Kolhatkar\affil{IPAC, California Institute of Technology 100-22, Pasadena, CA 91125}}
 
\begin{abstract}
We present a new phenomenological model for the spectral energy distribution of normal star-forming galaxies between 3 and 1100 $\mu$m.  A sequence of realistic galaxy spectra are constructed from a family of dust emission curves assuming a power law distribution of dust mass over a wide range of interstellar radiation fields.  For each interstellar radiation field heating intensity we combine emission curves for large and very small grains and aromatic feature carriers.  The model is constrained by \IRAS\ and ISOCAM broadband photometric and ISOPHOT spectrophotometric observations for our sample of 69 normal galaxies; the model reproduces well the empirical spectra and infrared color trends.  These model spectra allow us to determine the infrared energy budget for normal galaxies, and in particular to translate far-infrared fluxes into total (bolometric) infrared fluxes.  The 20 to 42 \m\ range appears to show the most significant growth in relative terms as the activity level increases, suggesting that the 20--42 \m\ continuum may be the best dust emission tracer of current star formation in galaxies.  The redshift dependence of infrared color-color diagrams and the far-infrared to radio correlation for galaxies are also explored.
\end{abstract}
 
 
\section {Introduction}
To better understand the implications of infrared and submillimeter observations of galaxies at high redshifts, it is first necessary to have a grasp of the emission characteristics of galaxies at low redshifts.  For example, comparisons between what is observed at high redshift and what is predicted by cosmological models buttressed by accurate spectral energy distribution (SED) templates of local galaxies can suggest how galaxies evolve with look-back time.  The infrared portion of galaxy SEDs has recently received particularly intense interest since the cosmic infrared background radiation is known to be twice as bright as the optical background, and the number counts in the mid-infrared are about an order of magnitude higher than predictions from zero evolution models (Pozzetti et al. 1998; Elbaz et al. 1999; Lagache et al. 1999).

A further motivation for constructing model infrared SEDs for galaxies is to quantify the infrared energy budget for various star-formation activity levels.  Success in this arena would prove especially timely---the initial exuberance over estimates of the cosmic star formation history has been tempered by the sobering realization that extinction corrections to star formation rates derived from visible/ultraviolet indicators are crucial, but assuredly much more difficult to measure.  Moreover, a normal galaxy SED model provides a standard by which to consider the role the mid-infrared aromatic features and other components play in the total infrared luminosity budget.  Finally, the ability to transform the canonical far-infrared ``FIR'' fluxes, which are scaled from the \IRAS\ 60 and 100 \m\ fluxes and nominally cover 42--122 \m\ (e.g. Helou et al. 1988), to 3--1100 \m\ total-infrared fluxes should prove useful, especially since the observed far-infrared flux may significantly differ from the rest frame far-infrared flux for higher redshift galaxies.  This can be an important transformation if far-infrared fluxes are used as a first-order indication of the star-formation activity level (e.g. Silva et al. 1998; Rowan-Robinson et al. 1997).   

SED models can be constructed from first principles and then adjusted accordingly to fit observations, they can be devised on a purely empirical basis, or a hybrid of these two methods can be used.  For example, Csabai et al. (1999) show that galaxy optical SEDs can be reliably reconstructed using only broadband fluxes (and redshifts for $k$-corrections); using their model they can reproduce Hubble Deep Field galaxy colors (UBVIJHK) to better than 10\% and can accurately predict their photometric redshifts.  The same approach can be taken for the development of infrared SEDs using \ISO\ and \IRAS\ data, providing yet another tool for studying galaxies at high redshifts.

There have been several other recent efforts aimed at developing galaxy infrared SEDs: Guiderdoni et al. (1998); Boselli et al. (1998); Siebenmorgan, Kr\"{u}gel \& Chini (1999); Devriendt, Guiderdoni \& Sadat (1999); Calzetti et al. (2000); Rowan-Robinson (2000); Granato et al. (2000); Garcia \& Espinosa (2000); Bianchi, Davies \& Alton (2000); Bekki \& Shioya (2000).  Many (but certainly not all) of these efforts focus on active galaxies such as Seyfert types or ultraluminous galaxies, and use a superposition of graybody curves at different temperatures to model the different components of the interstellar medium.  We present a new model for the infrared SED, one that builds on the pioneering work of D\'{e}sert, Boulanger \& Puget (1990, hereafter DBP90), improving their approach to very small grain emission and replacing their Polycyclic Aromatic Hydrocarbon (PAH) emission profiles with actual data from \ISO.  While DBP90 dealt solely with spectra of emission regions in the Milky Way, we synthesize their results with ours to construct galaxy-wide emission spectra.  An attractive feature of our model is its simplicity: the core assumption is a power-law distribution in the dust mass.  We empirically constrain the few parameters with observations of 69 normal galaxies. 

Section 2 describes our sample of galaxies, while Section 3 presents the corresponding data we use to constrain the model.  The empirical synthetic spectra are given in Section 4 and the model spectra and the underlying motivation for them are discussed in Section 5.  Some useful applications of the model are presented in Section 6.  A discussion of the results and our concluding remarks are contained in Sections 7 and 8.

\section {The Sample}

The data we use for constraining our SED model derive from the \ISO\ Key Project on the Interstellar Medium of Normal Galaxies (Helou et al. 1996; Dale et al. 2000) under NASA Guaranteed Time.  The parent sample for this project had the original following criteria: $f_\nu(60$ \m) $\gtrsim 3$ Jy; a published redshift; and no AGN or Seyfert classification on NED\footnote{The NASA/IPAC Extragalactic Database is operated by the Jet Propulsion Laboratory, California Institute of Technology, under contract with the National Aeronautics and Space Administration.} (at that time).  From this parent sample, Helou and collaborators collected multiwavelength data sets on a subsample that explores the full range of morphology, luminosity ($L_{\rm FIR}$ from less than $10^{8} L_\odot$ to as large as $10^{12} L_\odot$), infrared-to-blue ratio (0.05 to 50) and infrared colors (see Sections \ref{sec:iras-iras} and \ref{sec:iso-iras}) among star-forming galaxies.  While the sample is too small to perform any rigorous statistical analysis concerning its completeness, the sample does give a fair representation of the local population.  In terms of optical morphology, although all types are represented, the sample is largely comprised of S0, spiral, and irregular galaxies; only two galaxies are unambiguously classified as ellipticals.  And though the sample was selected to explore a large variety of normal star-forming galaxies and excluded ultraluminous and AGN-dominated galaxies, a few sources will inevitably show exceptional properties upon close inspection.  For example, the heavily obscured NGC 4418 is suspected to host an AGN (e.g. Roche et al. 1986) and although several studies suggest Mrk 331 is a normal star-forming galaxy (e.g. de Bruyn \& Wilson 1976; Fuentes-Williams \& Stocke 1988; Rush et al. 1996; Poggianti \& Wu 2000), NED now lists Mrk 331 as a possible Seyfert 2 galaxy.

Sixty of these objects were selected to be small in their \IRAS\ emission size compared to the 80$\arcsec$ ISOLWS beam and the 3$\arcmin$ ISOCAM field of view, to allow studies of their global properties.  In addition, nine nearby galaxies were mapped to the extent possible, including NGC 6946, NGC 1313, IC 10, and part of M 101.  For most galaxies, maps were obtained at 7 and 15 \m\ with ISOCAM, spectro-photometry was obtained with ISOPHOT-S between 3 and 12 \m, and far-infared fine structure lines were targeted with ISOLWS, attempting to measure as many as possible of the following lines, in the order listed: \CII\ $\lambda$157.7 \m, \OI\ $\lambda$63.2 \m, \NII\ $\lambda$121.9 \m, \OIII\ $\lambda$88.4 \m, \NIII\ $\lambda$57.3 \m, \OIII\ $\lambda$51.8 \m.  All Key Project galaxies were detected in all four \IRAS\ passbands: 12, 25, 60, and 100 \m.  The ISOCAM data are presented and analyzed in Dale et al. (2000); the ISOPHOT-S data are discussed in Helou et al. (2000) and Lu et al. (2000); the ISOLWS observations and analysis are found in Malhotra et al. (2000). 

\section {Empirical Evidence}

\subsection {\IRAS\ Color-Color Diagram}
\label{sec:iras-iras}

\begin{figure}[!ht]
\centerline{\psfig{figure=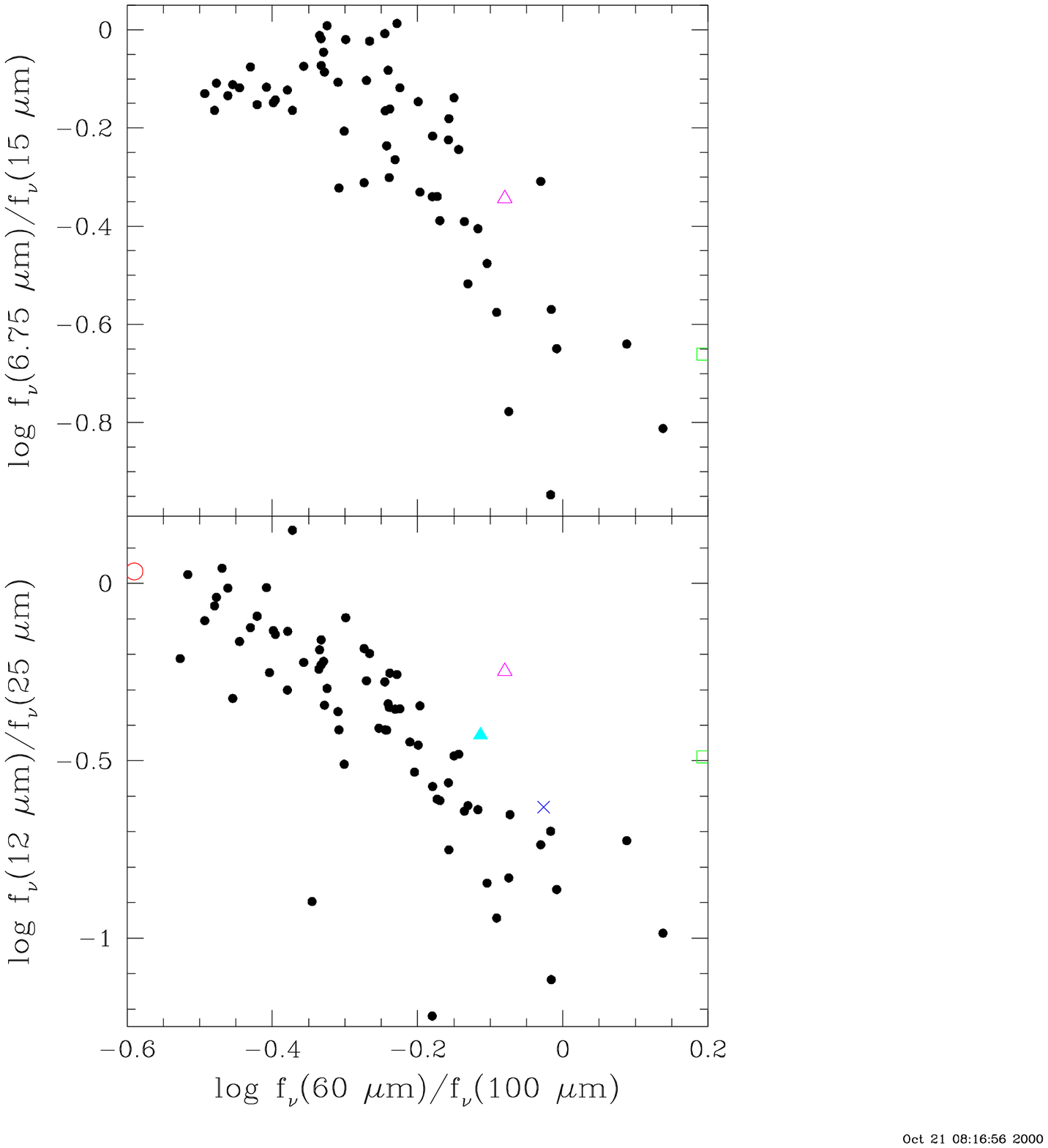,width=3.5in,bbllx=70pt,bblly=163pt,bburx=431pt,bbury=715pt}}
\caption[]
{\ The \IRAS\ and \ISO-\IRAS\ color-color diagrams for the \ISO\ Key Project sample of normal galaxies.  Included are the colors for a ``hyperluminous'' infrared galaxy (open square) and a radio quiet QSO (open triangle) derived from templates kindly provided by R. Cutri.  The HyperLIRG template is a composite SED from data for three such objects: \IRAS\ F15307+3252 (Cutri et al. 1994); \IRAS\ F10214+4724 (Rowan-Robinson et al. 1993 and Telesco 1993) and \IRAS\ P09104+4109 (Kleinmann et al. 1988).  They are normalized and combined to the approximate luminosity of F15307+3252 (see the lower panel of Figure \ref{fig:model}).  The radio quiet SED is taken from Elvis et al. (1994) and is a composite of a large number of observed radio quiet QSOs.  Also included for reference are the global \IRAS\ colors (from NED) for the quiescent galaxy M 81 (open circle), the starburst galaxy M 82 (cross), and a galaxy with an intermediate-level of star-formation activity: M100 (filled triangle).}
\label{fig:iso-iras}
\end{figure}

Data from the \IRAS\ mission provided a first important look into the gross characteristics of the mid-infrared and far-infrared sky.  One significant result is that normal star-forming galaxies follow a well-defined trend in \IRAScolorb\ vs \IRAScolor\ (Helou 1986).  Figure \ref{fig:iso-iras} displays such a diagram for our key project sample.  It turns out that this behavior can be explained as resulting from the interplay of two interstellar medium components with substantially different spectral properties: blackbody-like emission from classical grains in temperature equilibrium mostly at the longer wavelengths, and relatively fixed-shape mid-infrared emission from tiny grains composed of a few hundred atoms or less, intermittently heated by single photon events (Helou 1986).  

The sequence of infrared colors was clearly associated with a progression towards greater dust-heating intensity, as illustrated by the progression of colors in the California Nebula as one approaches the heating star (Boulanger et al. 1988).  The cool end of the color sequence corresponds to a cool, diffuse \HI\ medium and to quiescent molecular clouds, whereas the warm end corresponds to the colors of \HII\ regions, starbursts and galaxies with high $FIR/B$ ratios and higher infrared luminosity.  The warm end of the sequence is defined by greater values of \IRAScolor.  It is thus natural to associate the infrared color progression with a sequence of star formation activity in galaxies.

\subsection {\ISO-\IRAS\ Color-Color Diagram}
\label{sec:iso-iras}

The \ISO\ mission allowed a finer sampling over the span of mid-infrared wavelengths, with a total of 10 broad band filters available for doing photometry.  Most notable among these filters are those centered at 6.75 and 15 $\mu$m, the ``LW2'' and ``LW3'' filters.  Several large programs were carried out in these two bandpasses, including our own key project.  These filters were designed to capture mostly aromatic feature emission near 7 $\mu$m and the mid-infrared continuum beyond the bulk of the aromatic features at 15 $\mu$m, though the 12.7 $\mu$m feature does add a small amount to the LW3 emission.  It is thus not particularly surprising that the \ISOcolor\ ratio has emerged as an interesting diagnostic of the radiation environment.  It remains relatively constant and near unity as the interstellar medium of galaxies proceeds from quiescent to mildly active, where the level of activity is indicated by a rising \IRAScolor\ (Helou 1986; Dale et al. 2000).  As dust heating increases further, the flux at 15 \m\ increases steeply compared to 6.75 \m.  The data plotted in Figure \ref{fig:iso-iras} are consistent with an inflection in the mean trend occurring near log \IRAScolor =$-0.2$, which we interpret as more due to excess emission in the 15 \m\ band rather than mostly a drop in the 7 \m\ band flux.  The main argument for this interpretation is that the global $f_\nu(6.75 \mu {\rm m}) \over {\rm FIR}$ ratio does not drop as precipitously as \ISOcolor\ for these objects.  We assign to this mid-infrared emission a characteristic temperature 100 K $< T_{\rm MIR} <$ 200 K, since that is the range that would allow a blackbody to contribute considerably to the 15 \m\ band but not to the 7 \m\ band; the estimates hold for modified blackbodies as well.  Such values of $T_{\rm MIR}$ are typical of heating intensities about $10^4$ times greater than the diffuse interstellar radiation field in the Solar Neighborhood (Helou et al. 1997).  This color behavior is observed in our sample of galaxies (Figure \ref{fig:iso-iras}) as well as other samples (Vigroux et al. 1999).

While such a color temperature could result from classical dust heated within or just outside \HII\ regions, there is no decisive evidence as to the size of the grains involved.  It is simpler at this time to associate this component empirically with the observed emission spectrum of \HII\ regions and their immediate surroundings (Cesarsky et al. 1996; Tran 1998; Contursi et al. 2000).  This emission has  severely depressed aromatic feature or PAH emission, and is dominated by a steeply rising though not quite a blackbody continuum near 15 \m, consistent with mild fluctuations in grain temperatures of the order $\Delta T/T \sim 0.5$.  This \HII\ region hot dust component becomes detectable in systems where the color temperature from the \IRAScolor\ ratio is only $T_{\rm FIR} \approx 50$ K (Helou et al. 1988).  The disparity between color temperatures derived from different wavelengths demonstrates the broad distribution of physical dust temperatures within any galaxy.\footnote{The quoted far-infrared color temperatures are merely approximations arising from graybody profiles that exhibit similar flux ratios for the two wavelength bands.  See Helou et al. (1988) for more details on \IRAScolor\ color temperatures.}  The combined data from \ISO\ and \IRAS\ on these systems are consistent with an extension of the ``two-component model'' of infrared emission.  The low \ISOcolor\ ratio is associated with the active component, and combines in a variable proportion with a quiescent component where \ISOcolor\ is near unity (Dale et al. 1999).

\subsection {ISOPHOT Mid-Infrared Spectral Observations}
\label{sec:mir_spectrum}

\begin{figure}[!ht]
\centerline{\psfig{figure=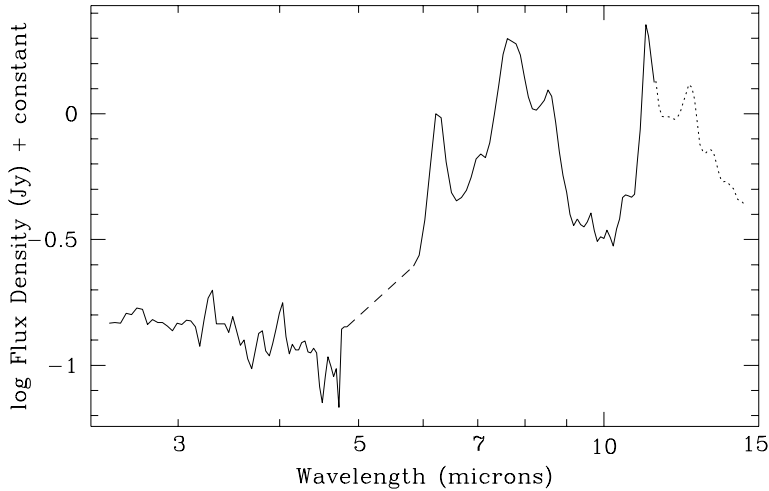,width=4in,bbllx=17pt,bblly=158pt,bburx=570pt,bbury=515pt}}
\caption[]
{\ A composite mid-infrared spectrum obtained from an average of ISOPHOT-S spectra for some 40 Key Project galaxies.  The ordinate is the flux density within the 24\arcsec $\times$24\arcsec\ ISOPHOT-S aperture, and the abscissa is the rest-frame wavelength.  The dashed line indicates the gap in the wavelength coverage, while the dotted extension at longer wavelengths represents our small modification to add the 12.7 \m\ aromatic feature (e.g. Cesarsky et al. 1996) since our ISOPHOT-S spectra do not extend to that wavelength.  See Lu et al. (2000) and Helou et al. (2000) for more details.}
\label{fig:mir_spectrum}
\end{figure}

In addition to broadband flux measures in the mid and far-infrared, we also have obtained mid-infrared spectra for a large portion of the Key Project sample.  As described in Helou et al. (2000) and Lu et al. (2000), ISOPHOT spectra were obtained for 43 galaxies over the wavelength ranges 2.5--5 \m\ and 5.7--11.6 \m\ (see Figure \ref{fig:mir_spectrum}).  These data are a direct indication of the typical behavior displayed by local normal galaxies at mid-infrared wavelengths.  The shapes and relative strengths of the features are quite similar to those seen in Galactic sources (nebulae, molecular clouds, diffuse atomic clouds, and \HII\ region surroundings; e.g. Cesarsky et al. 1996; Geballe 1997; Tokunaga 1997; Uchida, Sellgren \& Werner 1998) and a number of galaxies (e.g. Boulade et al. 1996; Vigroux et al. 1996; Metcalfe et al. 1996).  Helou and coworkers found that the relative fluxes of the spectral features (with the possible exception of the feature at 11.3 \m), and the general shape of the mid-infrared spectrum, vary little from galaxy to galaxy to within the 20\% ISOPHOT-S calibration uncertainty (Helou et al. 2000; Lu et al. 2000).  This aspect of the mid-infrared spectrum is observed for a wide range of parameters such as far-infrared colors and star formation activity level, strong evidence that the emitting particles are transiently excited by individual photons.

An important consequence of the invariant shape of the spectrum to 11 \m, even as the infrared-to-blue ratio reaches high values, is that the 10 \m\ trough is best interpreted as a gap between aromatic emission features rather than a strong silicate absorption feature.  An absorption feature would become more pronounced in galaxies with larger infrared-to-blue ratios, and that is not observed.  Using ISOSWS observations of starburst and Seyfert galaxies, reflection nebulae and star-forming regions, Sturm et al. (2000) show there is little evidence for significant 10 \m\ silicate absorption, a conclusion that is consistent with the interpretation of the above mid-infrared spectrum first presented by Helou et al. (2000).  This conclusion is reflected in our approach to modeling the spectra (\S \ref{sec:modeled_spectra}).

\section {Synthetic Empirical Infrared Spectra}

\begin{figure}[!ht]
\centerline{\psfig{figure=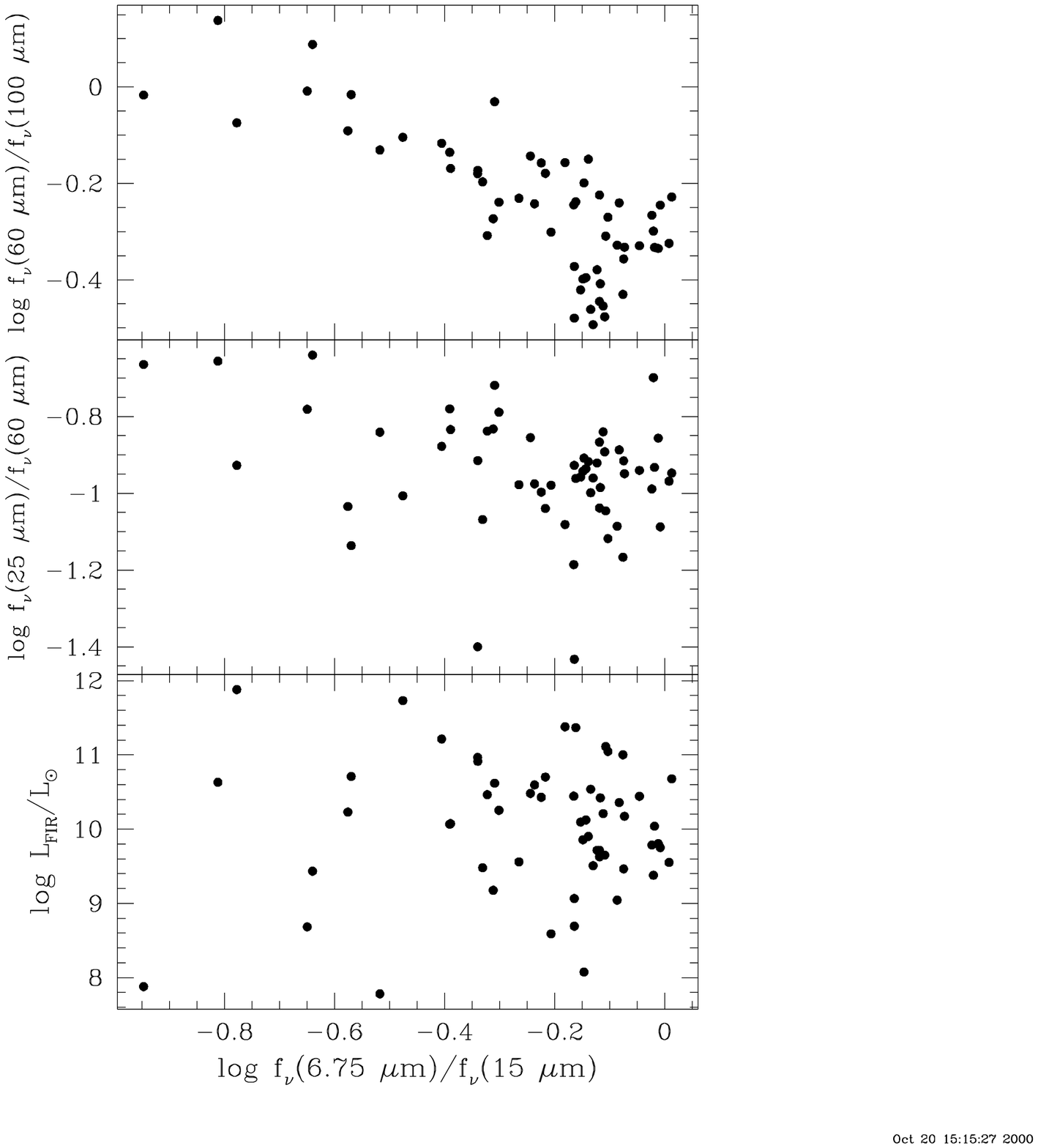,width=3.5in,bbllx=76pt,bblly=174pt,bburx=423pt,bbury=715pt}}
\caption[]
{\ A demonstration that, in contrast to the \IRAScolor\ ratio, the \IRAScolorc\ ratio and the far-infrared luminosity show relatively large scatter for a given \ISOcolor, a ratio known to track the relative intensity of the interstellar radiation field (e.g. Contursi et al. 1998).  Thus \IRAScolorc\ and $L_{\rm FIR}$ are not optimal parameters by which to develop a sequence in star-formation activity level.}
\label{fig:sort_alternatives}
\end{figure}

We construct synthetic empirical infrared spectra using \ISO\ and \IRAS\ broadband flux measurements and taking averages over classes of galaxies.  To do this effectively, it is important to first separate the galaxies into a number of representative classes according to some physical guideline.  It is customary to differentiate between galaxies according to luminosity, size, optical morphology, and other such standard parameters.  However, to construct infrared SEDs it seems more appropriate to use a criterion defined by the infrared properties of galaxies, and infrared colors and luminosity are logical choices.  As discussed in Section \ref{sec:iras-iras}, the ratio \IRAScolor\ is an indication of the general heating intensity of large grains and thus the overall level of star-formation activity within a normal galaxy.  Two possible alternatives are the far-infrared luminosity or the \IRAScolorc\ ratio.  The latter parameter may have no significance other than being an arbitrary ratio between the mid-infrared and far-infrared emission, though it may be linked to the destruction of very small grains and aromatics by intense star formation, suggesting that it signals an evolutionary stage parameter.  We find that \IRAScolor\ provides a superior sorting parameter, in the sense that the population dispersions in various infrared flux ratios are relatively small within a given class, or \IRAScolor\ `bin'.  A graphical indication of this superiority is reflected in the large scatter in the radiation field diagnostic \ISOcolor\ observed for a given \IRAScolorc\ or $L_{FIR}$ compared to that seen for a given \IRAScolor\ (Figure \ref{fig:sort_alternatives}).
In short, differentiating between galaxies according to their \IRAScolor\ ratio allows us to construct a rather tight sequence in global star-formation activity level.

We have chosen to separate the galaxies into seven bins of size 0.1 dex in \IRAScolor, a value somewhat larger than the typical uncertainty in that ratio: 0.06--0.09 dex for 10-15\% flux uncertainty.  The bin ranges and the number of galaxies within each bin are listed in columns 1 and 2 of Table \ref{tab:flux_ratios}.  The choice of exactly seven bins is somewhat arbitrary, but it does allow a useful compromise between ``resolution'' in far-infrared color and sufficient ``signal-to-noise'' within each bin.  Table \ref{tab:flux_ratios} also provides the average logarithmic flux ratios for all 15 \IRAS\ and \ISO\ broadband flux combinations for each bin.  The numbers given are computed from ratios of flux densities in Jy.  The numbers in parentheses are the r.m.s. variations in the ratios.

\begin{figure}[!ht]
\centerline{\psfig{figure=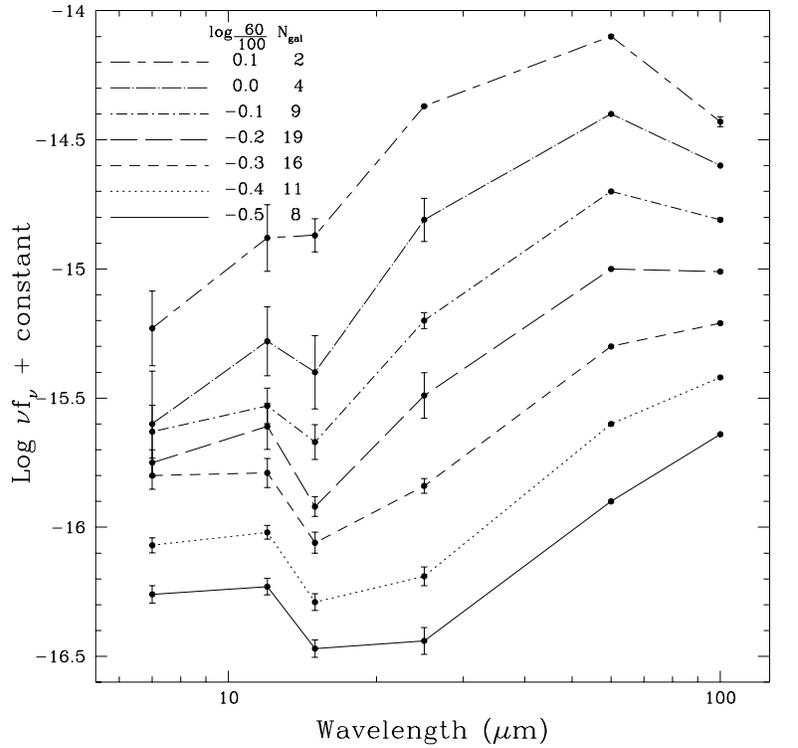,width=4in,bbllx=17pt,bblly=158pt,bburx=565pt,bbury=691pt}}
\caption[]
{\ The connected solid points in Figure \ref{fig:synthetic} indicate synthetic average galaxy empirical spectra for each \IRAScolor\ bin, determined via \IRAS\ and \ISO\ broad band flux measurements at 6.75, 12, 15, 25, 60, and 100 \m; the values displayed are the flux densities $f_\nu$ times the nominal bandpass frequencies, without any attempt at color correction.  We compute the displayed spectra by normalizing to the 60 $\mu$m flux and then averaging within the \IRAScolor\ bins.  The spectra are arbitrarily vertically shifted for spacing purposes.  Error bars stem from the 1$\sigma$ dispersions within each bin, divided by the square root of the number of objects within the bin; the error bars represent the bin spread in ${f_\nu (\lambda)} \over {f_\nu (60 \mu {\rm m})}$ for each $\lambda$.  There is no dispersion at 60 $\mu$m due to our normalization scheme, and little dispersion at 100 \m\ because of the sorting scheme.}
\label{fig:synthetic}
\end{figure}

There are some important systematic features to the spectra.  The far-infrared emission peaks at shorter wavelengths for galaxies with more intense heating environments (higher \IRAScolor).  This is widely attributed to the increased thermal heating of the classical large grain population, and the resulting change to its graybody emission profile.  The mid-infrared characteristics of normal galaxies also show systematic behavior: for the more quiescent galaxies, the flux at 6.75 \m\ is significantly greater than that at 15 \m.  The situation is reversed for the galaxies with more intense heating.  This trend is also linked to the strength of the heating environment and is equivalent to the effect discussed in the \ISO\ and \IRAS\ color-color diagrams (see Figure \ref{fig:iso-iras} and the corresponding discussion in Section \ref{sec:iras-iras} and Section \ref{sec:iso-iras}).  

\section {Modeled Spectra}
\label{sec:modeled_spectra}

In modeling the above spectra, we have adopted as a starting point the framework of DBP90, who used \IRAS\ observations of diffuse cirrus in the Solar Neighborhood to carefully energy-balance their model infrared spectra.  To construct the model SEDs we combine the emission curves from three dust components: large grains, very small grains, and the carriers of aromatic features (presumably PAHs).  The DBP90 models run from approximately 1 to 1000 $\mu$m in wavelength and span a heating intensity range of $U=0.3$ to 1000 in units of the local interstellar radiation field.  To test the validity of the model for a wide range of conditions, DBP90 also compared their model with observations of dark clouds, Galactic cirrus, reflection nebulae, and \HII\ regions.  We have updated their approach to very small grain emission using Draine \& Anderson (1985) temperature distribution profiles, and we have replaced their PAH emission spectrum with \ISO\ observations of the stable 3--12 \m\ mid-infrared spectrum.

\subsection{Large Grain Emission Profiles}
For the large grain emission curves we primarily rely on the formalism developed by DBP90 which uses a graybody emission (black body plus an assumed far-infrared emissivity law: $\epsilon \propto \nu^2$), the amplitude and wavelength peak systematically varying with $U$.  As outlined in DBP90, a power law model is used for the large grain size distribution, where the number density of grains with radius between $a$ and $a+da$ is:
\be
n(a) \propto a^{-2.9} \;\;\;\;\; {\rm with} \; a_{\rm min}=15 \; {\rm nm \; and} \; a_{\rm max}=110 \; {\rm nm}.
\ee
We use their large grain emission profiles without any modification.

\subsection{Very Small Grain Emission Profiles}
Some progress has been made in the understanding of the very small grain emission since the work of DBP90, though their existence is still questioned since we have yet to find an unambiguous characteristic spectral signature between 15 and 60 \m.  It is believed that the heating of the very small grains is intermediate between thermal equilibrium and single photon heating, resulting in mild thermal fluctuations $\Delta T/T < 1$.  We expect a near thermal graybody spectrum for very small grains bathed in extremely intense heating environments $U$, and conversely a broader emission profile at longer wavelengths for more quiescent environs, the governing parameter being the frequency with which heating photons are absorbed by each grain.  Draine \& Anderson (1985) have computed the temperature distribution profiles for a variety of grain sizes $a$ over a moderate range of incident heating intensities ($U$=[0.5,1,3]).  They find that graphite grains of size 0.02 \m\ in these environments radiate at essentially a single temperature near 20 K, whereas smaller graphite grains emit over a much larger temperature range of several hundred degrees Kelvin.  

\begin{figure}[!ht]
\centerline{\psfig{figure=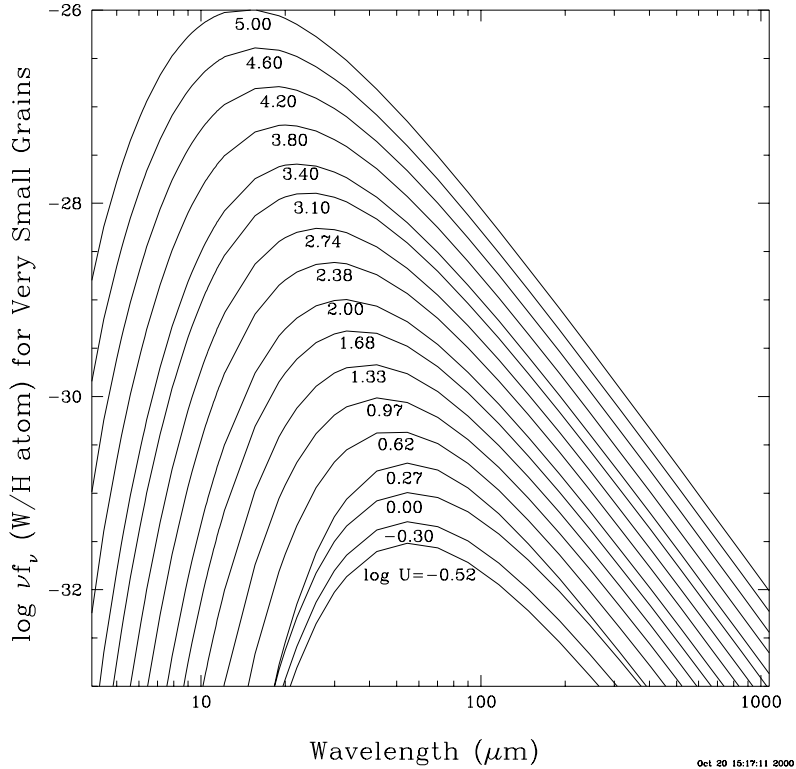,width=4in,bbllx=19pt,bblly=147pt,bburx=574pt,bbury=697pt}}
\caption[]
{\ A sampling of our collection of very small grain emission profiles for heating intensities ranging from 0.3 to 10$^5$ times the local interstellar radiation field.}
\label{fig:vsg}
\end{figure}

An interesting result from the work of Tran (1998) allows one to apply such temperature distributions for a much wider range of $U$: the set of temperature distribution profiles for a given grain size heated by a range of $U$ is conceptually equivalent to the set of profiles for a range of small grain sizes heated by a given $U$.  Tran (1998) quantified such a relationship by mapping $a$ to $U$ in this context.  We use this result and the work of Draine \& Anderson (1985) to construct very small grain emission profiles for $U$=0.3 to 10$^5$, with a diminishing departure from a true graybody profile for increasing values of $U$ (the assumed emissivity law is $\epsilon_\nu \propto \nu^2$).  The only respect of the DBP90 model we employ is the integrated flux normalization of the very small grain emission curves.  Figure \ref{fig:vsg} displays the very small grain emission profiles for the range $U=0.3$ to 10$^5$.

\subsection{PAH Emission Profiles}
Substantial progress has been made since 1990 in identifying and characterizing PAH emission, both from observations (e.g. Roelfsema et al. 1996; Verstraete et al. 1996) and work done in the lab (e.g. Hudgins \& Allamandola 1995, 1997).  For this component of the infrared spectrum, we spliced a slightly modified version of our observed normal galaxy mid-infrared spectrum (Helou et al. 2000; described in Section \ref{sec:mir_spectrum} and displayed in Figure \ref{fig:mir_spectrum}) into DBP90's PAH spectrum.  This component is set in amplitude by normalizing to the DBP90 model's flux within the \IRAS\ 12 \m\ bandpass.  We emphasize that it is appropriate to incorporate this spectrum into our normal galaxy SED model mainly because the spectrum stems directly from our own normal galaxy sample, and it has been shown that the shape and relative strengths of the features in the spectrum do not appreciably vary among normal galaxies (Helou et al. 2000; Lu et al. 2000).

\begin{figure}[!ht]
\centerline{\psfig{figure=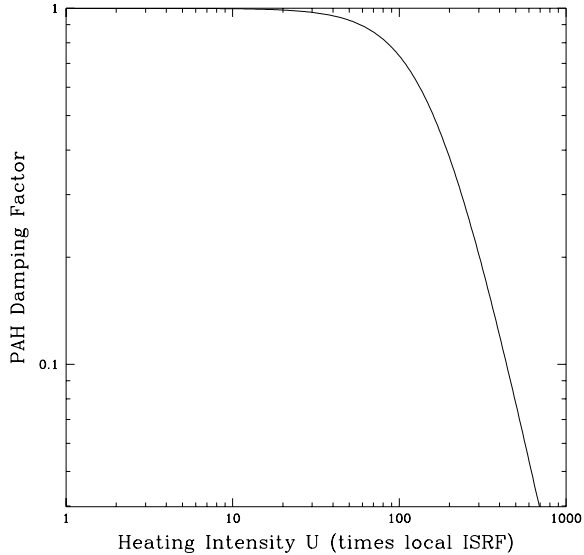,width=3in,bbllx=23pt,bblly=159pt,bburx=577pt,bbury=692pt}}
\caption[]
{\ The PAH damping factor as a function of $U$.  The curve follows a Fermi-Dirac profile, $(\exp(5\log[U/160])+1)^{-1}$, where the parameters have been fixed by minimizing the $\chi^2$ difference between the model and the empirical data (\S \ref{sec:comparison_with_obs}): the infrared color-color diagrams (Figure \ref{fig:iso-iras}) and the synthetic empirical spectra (Figure \ref{fig:synthetic}).}
\label{fig:pah_damp}
\end{figure}

There is a large body of evidence that indicates PAHs are likely to be destroyed, or that their emission is severely damped, in localized regions of high heating intensity such as the regions near OB associations (e.g. Boulanger et al. 1988; Cesarsky et al. 1996; Contursi et al. 1998; Lu et al. 2000).  Mid-infrared flux may also be diminished towards regions of high extinction in AGN-dominated galaxies, but it is unclear whether significant obscuration also occurs for starburst galaxies.  In fact, evidence has been presented that the shape of the starburst spectrum is unaffected by silicate absorption (Sturm et al. 1999).  Finally, low metallicity or a variable ratio of the different grain populations in active regions may suppress PAH emission.  In any event, to better match the empirical trends we implement a damping factor that monotonically decreases the integrated flux from the DBP90 PAH emission for models with increasing $U$.  This factor is displayed in Figure \ref{fig:pah_damp} as a function of $U$.

\subsection{Composite Model Spectra}
\begin{figure}[!ht]
\centerline{\psfig{figure=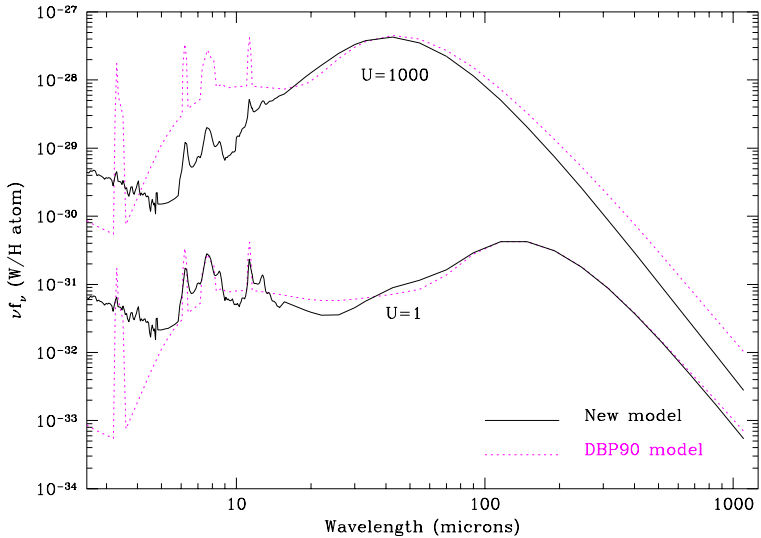,width=4in,bbllx=17pt,bblly=196pt,bburx=577pt,bbury=590pt}}
\caption[]
{\ A comparison of our model (solid lines) with the prior work of DBP90 (dotted lines).  The upper two lines correspond to a heating intensity of $U=1000$, whereas the lower two lines reflect the models' prediction for $U=1$.  The large discrepancy between the $U=1000$ models from 5 to 15 \m\ is due to our implementation of increased PAH destruction for regions of increased heating intensity.}
\label{fig:model-desert}
\end{figure}

Figure \ref{fig:model-desert} displays a sampling of model spectra composed of a superposition of large grain, very small grain, and PAH emission curves.  We adopt the overall normalization of DBP90, before PAH damping, to scale the relative contributions of the three grain types: $10^4 m_{\rm grain}/m_H=4.3, 4.7, 64$ for PAH, very small grains, and large grains, respectively.  The solid lines reflect our model's prediction for a dust mass in a heating environment where $U$ equals 1 and 1000, while the dotted lines do the same for DBP90.
The most obvious difference in the models is the discrepancy from 5 to 15 \m\ for $U=1000$, which highlights our PAH damping scheme for regions of increased heating intensity.  At longer wavelengths, notice that the DBP90 model shows an enhanced $U=1000$ emission.  Though at first glance this difference may suggest our lack of an extremely cold dust component, the difference actually arises from more significant far-infrared contributions for the DBP90 very small grain model.

\subsection{Final Model Spectra}
At this stage we have synthetic SEDs appropriate for a range of dusty environments bathed in different radiation fields.  We will hereafter refer to these as ``local'' SEDs.  To compute a galaxy SED, we rely on the fact that a galaxy is conceptually composed of a superposition of \HII\ regions, photodissociation regions and molecular clouds, cirrus-dominated environments, etc.  We combine the local SEDs assuming a power-law distribution in a given galaxy of dust mass over heating intensity:
\begin{eqnarray}
\label{eq:dMdU}
             dM_d(U) &\propto& U^{-\alpha} \; dU,\;\;\; 0.3 \leq U \leq 10^5
\end{eqnarray}
where $M_{\rm d}(U)$ is the dust mass heated by a radiation field at intensity $U$, and the exponent $\alpha$ is a parameter that represents the relative contributions of the different local SEDs.
The choice of a power law representation for $M_{\rm d}(U)$ is justified by the following general arguments which apply at two extremes of the heating conditions.  In the case of a diffuse medium where the heating intensity falls off primarily because of $r^{-2}$ dimming as one moves away from the heating source, one would write $dU/dr \propto - r^{-3}$.  For a uniform medium with $dM_{\rm d} \propto r^2dr$, one gets $dM_{\rm d} \propto U^{-2.5}dU$.  In the other scenario, the case of a dense medium where the heating intensity is primarily attenuated by dust absorption, one would write $dU/dr \propto -U$ for a slab.  For a uniform medium, $dM_{\rm d} \propto dr$, so one gets $dM_{\rm d} \propto U^{-1} dU$.  The first case would be a reasonable approximation for the diffuse cirrus-like components of the interstellar medium, whereas the second case would be a good approximation for a photodissociation region near young stars.  Intermediate cases between the two given here should yield intermediate values for the exponent in the power law scaling ($1 \lesssim \alpha \lesssim 2.5$).  In other words, a variable $\alpha$ parameter represents the varying contributions of active and quiescent regions from galaxy to galaxy.

\begin{figure}[!ht]
\centerline{\psfig{figure=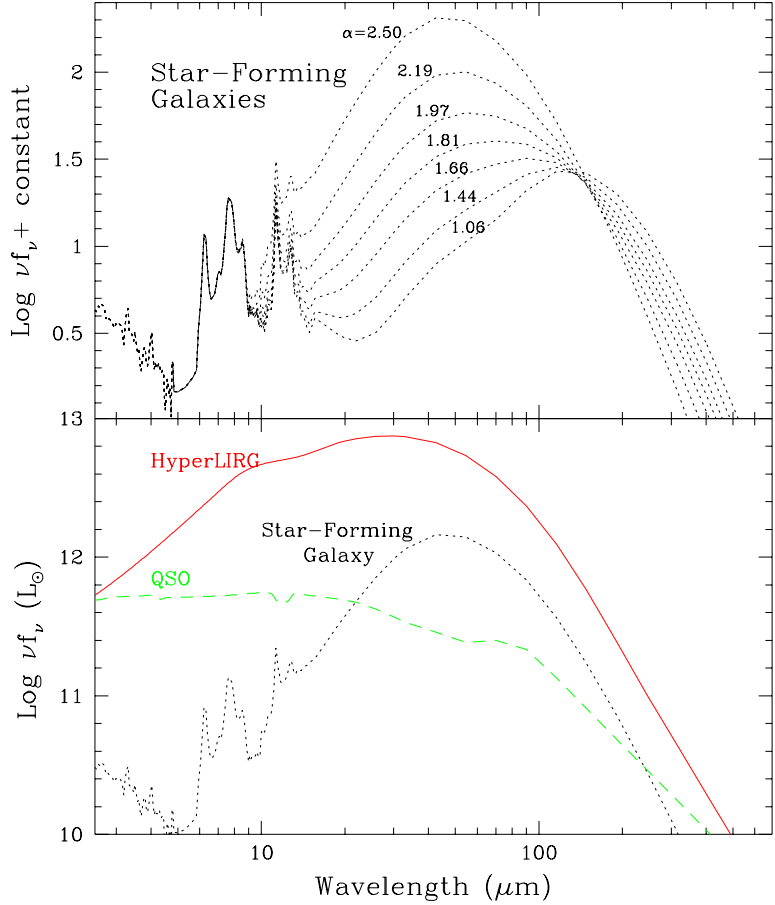,width=4in,bbllx=27pt,bblly=163pt,bburx=493pt,bbury=715pt}}
\caption[]
{\ Top: A sampling of the galaxy model spectra, arbitrarily normalized at the 6.2 \m\ feature.  These seven spectra have far-infrared colors similar to those of the seven average empirical spectra displayed in Figure \ref{fig:synthetic}. Bottom: A comparison between our model for an actively star-forming normal galaxy and SEDs for a hyperluminous infrared galaxy and a radio quiet QSO (see Figure \ref{fig:iso-iras}).  The normal galaxy here is the topmost curve in the upper panel normalized to have $L_{\rm FIR}=10^{12} L_\odot$, typical for galaxies in the \IRAS\ 1.2 Jy survey with log\IRAScolor $\simeq0.1$.}
\label{fig:model}
\end{figure}

The upper panel of Figure \ref{fig:model} displays seven sample model galaxy spectra\footnote{The progression of typical FIR luminosities for \IRAS\ 1.2 Jy survey galaxies with similar far-infrared colors is: $\log L_{\rm FIR}=8.8,9.8,10.4,11.0,11.4,11.8,12.1 L_\odot$.  Be aware, however, that \IRAScolor\ vs $\log L_{\rm FIR}$ shows {\it considerable} scatter.} which approximately match the seven empirical ones displayed in Figure \ref{fig:synthetic}.  
The progression of the far-infrared peak towards lower wavelengths for increased global heating intensities is obvious, as is the diminishing role played by the aromatic features in the overall infrared luminosity.  The former effect results from the shifting graybody profiles of the large and very small grains, while the latter effect is partly due to our enforced mode of PAH destruction.

\subsection{Comparison with Observations}
\label{sec:comparison_with_obs}
\begin{figure}[!ht]
\centerline{\psfig{figure=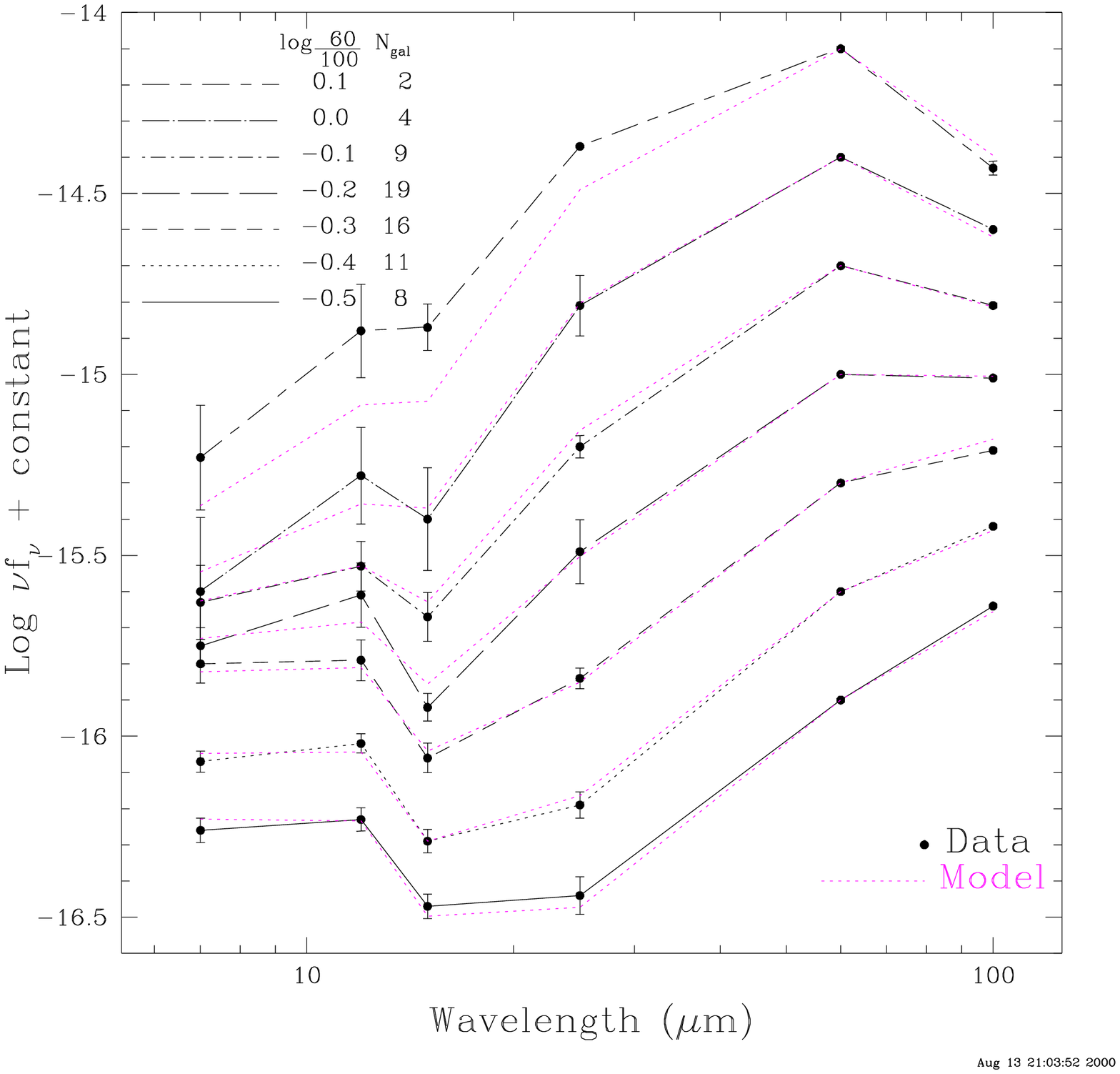,width=4in,bbllx=17pt,bblly=158pt,bburx=565pt,bbury=691pt}}
\caption[]
{\ A comparison of the model with the synthesized empirical galaxy spectra.  Model broadband fluxes are computed using our model SEDs and the \IRAS\ and ISOCAM filter profiles.}
\label{fig:model-synthetic}
\end{figure}

These model galaxy spectra agree quite well with the empirical synthetic spectra (see Figure \ref{fig:model-synthetic}).  
To execute this comparison, we convolved the \ISO\ and \IRAS\ broadband filter bandpasses with the model spectra; the known filter profiles and widths enabled us to plot equivalent quantities for the empirical and model spectra in Figure \ref{fig:model-synthetic}.  Here we have selected the seven model spectra that best match the sequence of synthetic empirical spectra (i.e. via chi-squared minimization), and normalized the model to observations at 60 \m.  The agreement is relatively poor for the highest \IRAScolor\ color because our sample includes only two galaxies within this bin.  Our model also predicts systematically low 12 \m\ emission with the discrepancy larger for the more active galaxies, a likely signal that we may have missed a latent parameter in the development of our mid-infrared emission curves.

If the range of local SEDs is limited to $U=0.3-10^5$, as we have done here, our sample of normal galaxy synthetic empirical spectra (Figure \ref{fig:synthetic}) span a range of $\alpha=1$ to 2.5, with $\alpha$ values near unity representative of more active galaxies.  This is a remarkable agreement with the simple scaling arguments offered to justify the use of power laws, especially with the coldest galaxies being fit with $\alpha=2.5$ as estimated for cirrus-like media, and the warmest galaxies being fit with $\alpha=1$ as estimated for photodissociation regions in star forming regions.

\begin{figure}[!ht]
\centerline{\psfig{figure=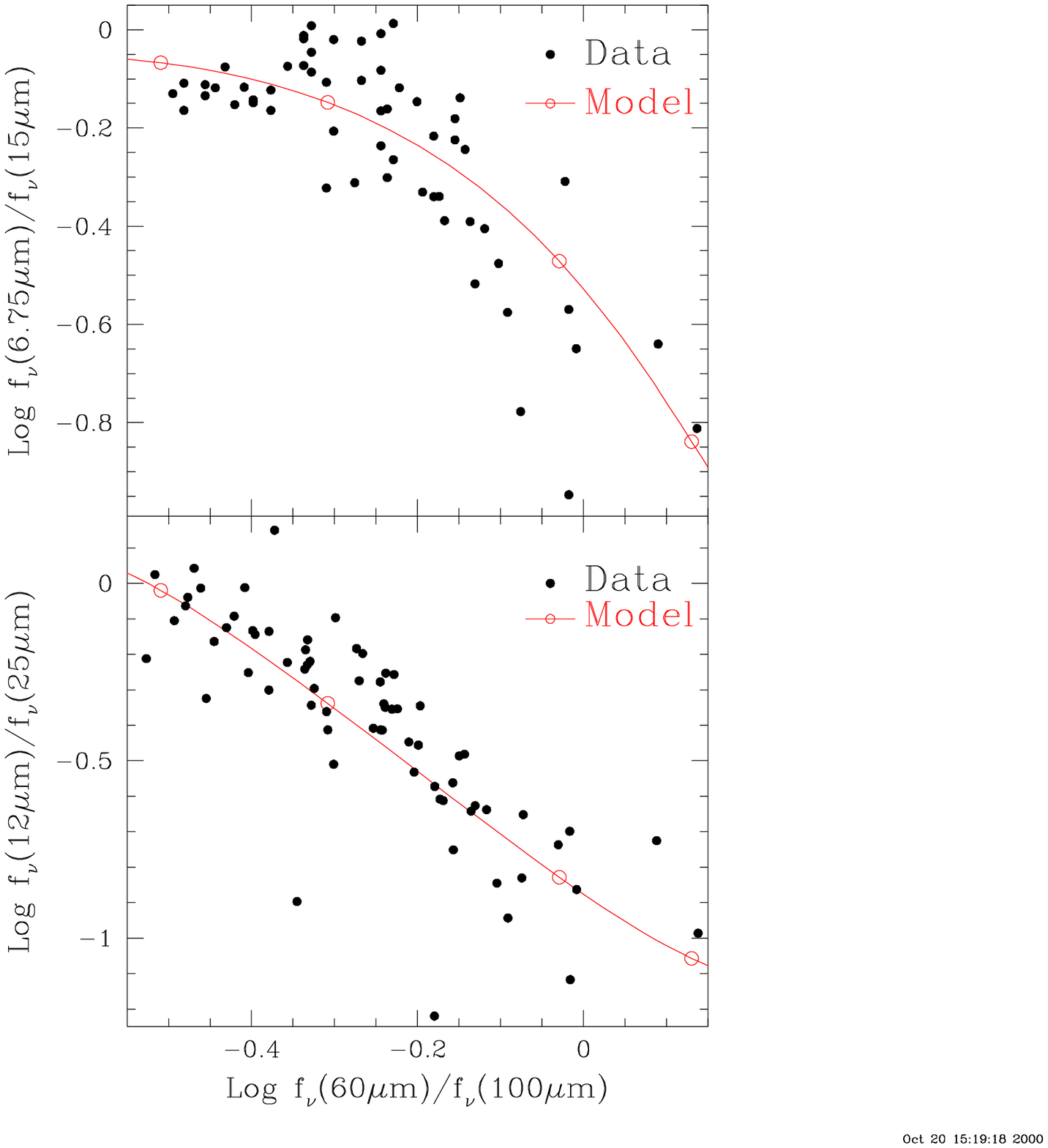,width=3.5in,bbllx=70pt,bblly=163pt,bburx=425pt,bbury=715pt}}
\caption[]
{\ A comparison of the model with the \IRAS\ and \ISO\ color-color diagrams.  The sequence of open circles, starting from the upper lefthand corner in each panel, indicate model colors that correspond to power law exponent values of $\alpha=$[2.5,2.0,1.5,1.0].}
\label{fig:color-color-model}
\end{figure}

Figure \ref{fig:color-color-model} shows the run of \ISO\ and \IRAS\ infrared colors produced by the model overlayed on our empirical galaxy data.  It should be noted that the run of colors for the Vigroux et al. (1999) galaxy sample also agrees quite well with the predictions of our model.

\section {Applications}

\subsection{Infrared Energy Budget}
A proper determination of the total infrared luminosity and the long wavelength spectral shape for normal galaxies is critical to estimating the galactic contribution to the infrared and submillimeter backgrounds, and thereby deriving the infrared term in the star formation history of the Universe.  An important first step towards meeting these goals can come from an accurate infrared energy budget for normal galaxies, such as the one presented In Table \ref{tab:IR_budget}.  The table shows how much energy emerges in various infrared bands for model galaxies with different star formation activity, parametrized by the \IRAS\ \IRAScolor\ color in the second column.  The headings for columns 3--8 give the wavelength range over which the spectrum is integrated, and the table entries are the fraction of total infrared luminosity appearing in that range (columns 9 and 10, on the other hand, show the ratio of PAH and very small grain emission relative to the large grain emission, integrated over 3--1100 \m).  The spectral range in column 7 corresponds to the ``FIR'' synthetic band (Helou et al. 1988 and Section \ref{sec:bolom}).  This particular wavelength range accounts for about half of the total infrared emission for galaxies with \IRAScolor\ $\gtrsim0.5$, but quickly drops by 20\% for cooler galaxies.  We remind the reader that these numbers reflect the average properties within each \IRAScolor\ bin; they ignore variations in spectral shape at constant \IRAScolor, including intrinsic scatter in the ratio of mid-infrared to far-infrared emission such as the \IRAScolorc\ or ${f_\nu (6.75 \mu {\rm m})} \over {f_\nu (100 \mu {\rm m})}$ ratios (Lu et al. 2000).

An important factor to be gleaned from these numbers is the relative contributions at mid-infrared and far-infrared wavelengths, or for a quantitative example, the ratio of the emission from 5--13 \m\ and 42--122 \m.  Given the observed trends in the infrared color-color diagrams, it is not surprising that this ratio drops rapidly as the level of star-formation activity increases (as indicated by a rising \IRAScolor).  This follows the trend already noted in Helou, Ryter \& Soifer (1991); Boselli et al. (1997, 1998) have interpreted similar trends as evidence for the destruction of the carriers of aromatic features in more intense radiation fields.  The 20 to 42 \m\ range appears to show the most significant growth in relative terms as the activity level increases, at the expense of the emission at submillimeter wavelengths, suggesting that the 20--42 \m\ continuum may be the best dust emission tracer of current star formation in galaxies.  This trend reflects the increased contributions from very small grains at shorter wavelengths for more intense heating environments.

\subsection{Total vs Far-Infrared Flux}
\label{sec:bolom}
The far-infrared flux, defined as the flux between 42 and 122 \m, is a favorite infrared diagnostic.  It is the most commonly quoted infrared flux, and quite often it is used as an indicator of the total level of activity in the interstellar medium.  Helou et al. (1988) provide a useful way of deriving the far-infrared flux from \IRAS\ measurements: FIR (W m$^{-2})=1.26 \cdot 10^{-14}(2.58 f_\nu(60 \mu{\rm m})+f_\nu(100 \mu{\rm m}))$, where $f_\nu(60 \mu{\rm m})$ and $f_\nu(100 \mu{\rm m})$ are in Jy.  This method is good to better than 1\% for graybody emission with temperatures between 20 and 80 K and for an emissivity between $\nu^0$ and $\nu^2$.  Helou and collaborators argue that this property should therefore also apply to realistic spectral energy distributions of galaxies, because those must be made up of a superposition of modified blackbodies in this temperature range.  Using our model SED curves, we have confirmed the 1\% accuracy of the method.  It is essentially a coincidental result of the properties of the \IRAS\ filter shapes that the simple linear combination allows us to estimate the luminosity in a well defined spectral window.  The real interest of far-infrared luminosity, however, is that this window encompasses a large fraction, and therefore a representative measure, of the total infrared luminosity.  

\begin{figure}[!ht]
\centerline{\psfig{figure=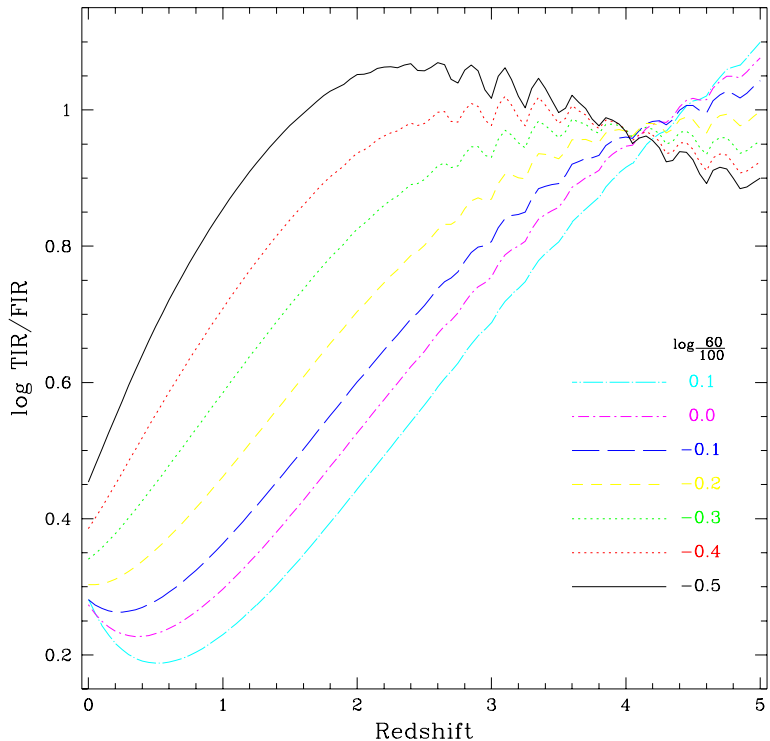,width=4in,bbllx=25pt,bblly=168pt,bburx=564pt,bbury=688pt}}
\caption[]
{\ The ratio of the total (3--1100 \m) to far-infrared (42--122 \m) flux.  The different lines correspond to the bolometric correction for spectra with rest frame far-infrared colors similar to those of the seven average empirical spectra displayed in Figure \ref{fig:synthetic}.}
\label{fig:bolom}
\end{figure}

With our SED model we can derive bolometric corrections for the purpose of converting observed far-infrared fluxes to the total infrared luminosity and therefore the dust mass.  The relation is displayed in Figure \ref{fig:bolom} as a function of redshift for spectra with a variety of intrinsic (rest frame) \IRAScolor\ ratios.     
At zero redshift the correction is only a factor of two to three, whereas it rises to one order of magnitude for higher redshifts.  The small scale features in the bolometric trends with redshift, caused by the influence of the mid-infrared features as they pass through the FIR spectral range, are smaller in amplitude for the more active galaxies since the mid-infrared features in these galaxies are much lower in flux than the far-infrared peak.  

The increase in the bolometric correction as a function of redshift occurs more quickly for the more quiescent galaxies, as the far-infrared peak for these sources lies beyond 100 \m\ at zero redshift and moves to even longer wavelengths for higher redshifts.  In contrast, the bolometric correction factor for the most active normal galaxies does not immediately rise with redshift, and in fact decreases a bit until bottoming out near $z \sim 0.6$, because the far-infrared peak is near 45 \m\ at zero redshift, and moves progressively farther into the FIR band as we proceed to intermediate redshifts.  In other words, the spread in the infrared bolometric correction factor at a given redshift is a function of the different SED shapes for galaxies with different far-infrared colors and thus global star-formation activity levels.  Moreover, this spread rapidly increases with redshift, rising to a maximum of a factor of four at a redshift of 1--2, and then decreases until actually reversing near $z=4$, beyond which the more active galaxies show larger infrared bolometric corrections.  Of course uncertainties in our SED model translate to uncertain bolometric corrections, but the general trends observed in Figure \ref{fig:bolom} should be kept in mind when planning high redshift observations with the Space Infrared Telescope Facility (\SIRTF) and other future infrared platforms. 

We have also computed the infrared bolometric corrections, for a given redshift, as a function of the observed (redshifted) \IRAScolor\ ratio.  The zero redshift trend is:
\be
\log \left({\rm TIR} \over {\rm FIR}\right) = a_0 + a_1 x + a_2 x^2 + a_3 x^3 + a_4 x^4
\label{eq:TIR_FIR}
\ee
where $x=\log $\IRAScolor\ and [$a(z$=$0)$]$\approx$[0.2738, $-$0.0282, 0.7281, 0.6208, 0.9118].  It is worth comparing our results to published bolometric corrections.  Using archival \IRAS\ and new 150 and 205 \m\ \ISO\ photometry to model the $\lambda \gtrsim40$ \m\ dust emission for a sample of six nearby ($z<0.03$) starburst galaxies, Calzetti et al. (2000) claim the total infrared (1--1000 \m) bolometric correction is log ${{\rm TIR} \over {\rm FIR}}=0.24$, some 15-40\% smaller than what we find for galaxies with similar \IRAScolor\ at zero redshift.  Sanders \& Mirabel (1996) also give total infrared (8-1000 \m) bolometric calculations for luminous and ultraluminous infrared galaxies, in this case as a function of the four \IRAS\ fluxes.  For the active end of our sequence of star-forming galaxies, the agreement with the Sanders \& Mirabel result is good to 4\%.  The agreement is much worse for more quiescent systems, with the Sanders \& Mirabel total infrared fluxes about 25\% lower than our models predict.  These discrepancies may be expected, in the sense that ultraluminous infrared galaxies are more actively star-forming than normal galaxies and thus should have a larger fraction of their dust emission at shorter wavelengths.  The discrepancies may also simply be due to differences in methodology: Sanders \& Mirabel compute their total infrared flux by fitting a single temperature dust emissivity model ($\epsilon \propto \nu^{-1}$) to the four \IRAS\ fluxes.

\subsection{Infrared Color-Color Diagrams at Higher Redshifts}
\begin{figure}[!ht]
\centerline{\psfig{figure=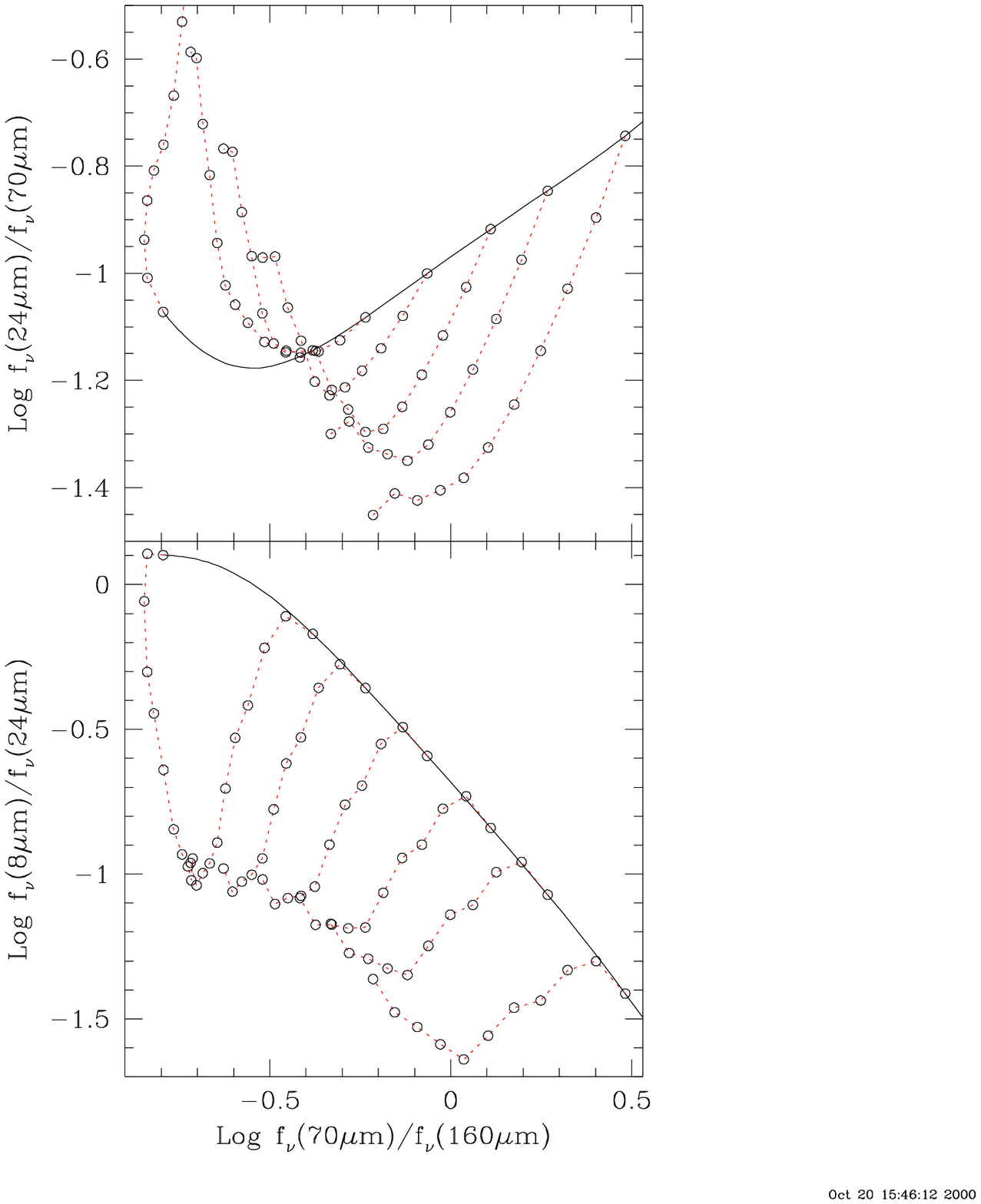,width=3.5in,bbllx=128pt,bblly=167pt,bburx=435pt,bbury=714pt}}
\caption[]
{\ Predicted \SIRTF\ color-color diagrams.  The smooth solid line indicates the zero redshift trend defined by the sequence of normal star-forming galaxies; the dotted lines are redshift tracks (from $z=0$ to 1) for the seven representative normal galaxies of Figure \ref{fig:model}.}
\label{fig:sirtf-sirtf-z}
\end{figure}

Optical color-color diagrams have recently reemerged as a powerful tool for yielding rough photometric redshifts of high redshift galaxies (e.g. Steidel et al. 1996).  Figure \ref{fig:sirtf-sirtf-z} shows predictions for the redshift dependence up to $z=1$ of infrared color-color diagrams using \SIRTF\ filter bandpasses.
The interesting wiggles in the redshift tracks reflect the redshift-dependent role of the aromatic emission features as they pass through the mid-infrared bandpasses.  The redshift tracks in the lefthand panel would be difficult to use as a diagnostic tool, as the curves begin to cluster together for increasing $z$.  The righthand color-color diagram, on the other hand, shows a cleaner separation between different galaxy types at $z=1$.

\subsection{Infrared to Radio Ratio as a Function of Redshift}
\begin{figure}[!ht]
\centerline{\psfig{figure=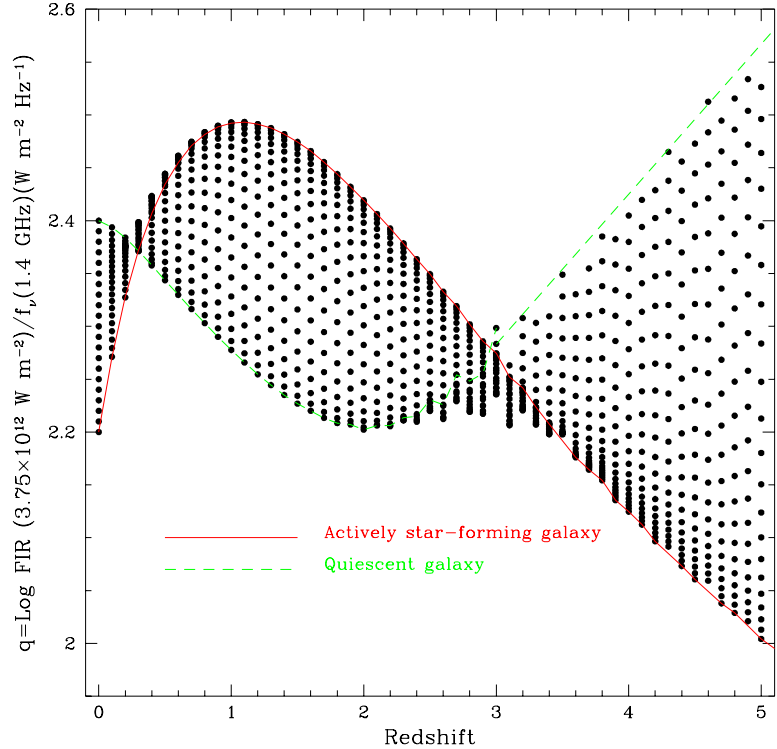,width=4in,bbllx=29pt,bblly=172pt,bburx=565pt,bbury=693pt}}
\caption[]
{\ Far-infrared to radio relation as a function of redshift.  The spread in $q$ at zero redshift is sized to reflect the 1$\sigma$ scatter in the ratio for nearby galaxies, and for illustrative purposes we have assigned the more quiescent (active) systems to exhibit ratios higher (lower) than the median trend; we have tracked the redshift dependence of the 1$\sigma$ envelope by mapping the $\langle q \rangle + 1\sigma$ ratio to our $\alpha=2.31$ model and the $\langle q \rangle - 1\sigma$ ratio to our $\alpha=1.06$ model.  Though this mapping is artificial, there is evidence for this type of behavior from observations that galaxies with relatively low far-infrared luminosities have on average even lower radio luminosities, as suggested by the analysis of Condon, Anderson \& Helou (1991).}
\label{fig:fir_to_radio}
\end{figure}

The far-infrared luminosity and radio continuum flux density for normal galaxies have been known to be correlated, and their ratio to have a remarkably small scatter, over a large range of galaxy types and luminosities (see Condon 1992 for a review).  The logarithmic measure 
\be
q = \log\left( {{\rm FIR} \over 3.75 \times 10^{12} {\rm W m}^{-2}} \right) - \log \left( {f_\nu \over {\rm W m}^{-2} {\rm Hz}^{-1}} \right).
\ee
defined by Helou, Soifer \& Rowan-Robinson (1985) has a median value of $\langle q \rangle \sim 2.3$ at $\nu \sim 1.4$ GHz and a 1$\sigma$ scatter of 0.2.  However, this ratio and its tightness will change with redshift due to $k$-corrections on both FIR and $f_\nu(1.4{\rm GHz})$.  Figure \ref{fig:fir_to_radio} shows this effect.  
Here we have assumed no evolution in the spectral shapes (including a fixed radio continuum slope of $f_\nu(1.4{\rm GHz}) \propto \nu^{-0.8}$) and that background effects are insignificant for 20 cm measurements, even though the cosmic microwave background radiation energy density increases as $(1+z)^4$ and may ``quench the radio emission at redshifts $z\gtrsim1$'' (Condon 1992).  Though the median value does not significantly change with redshift, the envelope of the 1$\sigma$ scatter shows an interesting dependence on $z$.  The 1$\sigma$ scatter is dramatically reduced near redshifts 0.4 and 3, with the more active systems showing higher ratios than the less active galaxies between these redshifts, contrary to the zero-redshift population.

\section {Discussion}

Models of the infrared spectral energy distribution of normal star-forming galaxies allow us to interpret and quantitatively assess the relative significance of various mid-infrared and far-infrared bands.  For example, we have seen that the 20 to 42 \m\ wavelength range exhibits the largest growth (a factor of five) in the infrared energy budget as the average heating intensity of the interstellar medium increases.  This strong trend reflects the increasingly important role of the very small grains, and to a lesser extent the large grain population, in regions of higher heating intensity, and is a clear signal that the 20 to 42 \m\ wavelength range is a useful indicator of star-formation activity level.  This is in contrast to the more standard 42 to 122 \m\ far-infrared wavelength range, for which the flux is often used as a star-formation standard (e.g. Silva et al. 1998).  This wavelength range, along with the 13 to 20 \m\ regime, only shows a fractional increase of 50\% with respect to the total infrared emission and should therefore be primarily used to estimate the total infrared luminosity, regardless of its origin in OB or older stars. 

On the other hand, for every percentage increase in the infrared energy budget by the 20 to 42 \m\ wavelength range there must be a corresponding decrease at other wavelengths.  From the distribution laid out in Table \ref{tab:IR_budget} it appears that the bulk of the contribution comes from the 122 to 1100 \m\ submillimeter wavelength range.  This regime falls from comprising 40\% of the total infrared emission for inactive, cirrus-dominated (`cold') galaxies to only 5\% for the galaxies with the most intense global interstellar heating. It thus appears that blind high redshift searches for sources at submillimeter wavelengths may yield a significant fraction of cold (and ultraluminous) galaxies, in contrast to conventional far-infrared surveys which preferentially pick up more active galaxies at higher redshifts.

At shorter wavelengths, our results echo those of Helou et al. (2000).  The fraction of starlight processed through the smallest grains and PAHs has been debated since the days of the \IRAS\ mission (e.g. Helou, Ryter \& Soifer 1991).  We estimate that the dust emission from 3 to 13 \m, a wavelength span for which the flux is primarily dominated by aromatic features and the underlying continuum, accounts for 5\% to 16\% of the total infrared dust luminosity from 3 to 1100 \m.  If we differentiate according to grain type in our model, we see that PAHs (very small grains) are responsible for between 4\% and 21\% (17\% and 22\%) of the total 3 to 1100 \m\ emission, the exact percentage depending on the activity level of the interstellar medium.

Which wavelength range best scales with the infrared bolometric luminosity?  The answer depends on the redshift involved.  Table \ref{tab:IR_budget} indicates that none of the broad bands has a fraction of the bolometric luminosity that is roughly the same for all galaxy types.  However, at $z=1$ the 20 to 42 \m\ wavelength range shows a flux that remains a stable $\sim 7$\% of the total infrared luminosity, and at $z=2$ the 122 to 1100 \m\ wavelength range is responsible for about 60\% of the total flux, independent of the global star-formation activity level.  From Figure \ref{fig:bolom} we see that the 42 to 122 \m\ wavelength range contributes a fairly constant 10\% of the total flux at $z=4$.

There are limitations to our method of modeling the infrared spectral energy distribution of normal star-forming galaxies.  \\
$\bullet$ First, we have not explicitly included a term that represents extremely cold dust residing in dark molecular clouds ($T<15$ K).  Such a term would obviously influence the far-infrared spectral shape of our models for the least active galaxies, and their submillimeter emission may thus be underestimated.  \\
$\bullet$ Second, the location of the turnover in our trend of PAH emission damping as a function of the interstellar radiation field occurs at too low an intensity (Figure \ref{fig:pah_damp}).  From spatially resolved work on Galactic and Magellanic \HII\ regions we know that the mid-infrared aromatic features begin to disappear and the \ISOcolor\ ratio starts to decrease only for local heating intensities above $10^3$ times the Solar Neighborhood value (Cesarsky et al. 1996; Contursi et al. 1998).  \\
$\bullet$ We assume optically thin infrared emission, though this may not strictly hold for regions in the most actively star-forming galaxies.  \\
$\bullet$ Another limitation of our model is that the maximum heating intensity in the interstellar medium is not exactly $U_{\rm max}=10^5$, but rather is expected to vary from galaxy to galaxy.  Nor do we expect the minimum value to remain a constant $U_{\rm min}=0.3$, but this concern is reflected in our first point about the lack of cold dust in our model.  \\
$\bullet$ A final caveat is the possible over-simplicity of our model.  The data in Figure \ref{fig:color-color-model} show that real galaxy flux ratios scatter about the model by up to a factor of two.  Perhaps another parameter such as the \IRAScolorc\ flux ratio, the PAH-to-large grain ratio, or the very small grain-to-large grain ratio can provide the necessary elasticity to better model the observed variations in star-forming galaxies.

\section {Conclusions}
Using \IRAS\ and \ISO\ data for our sample of 69 nearby galaxies as constraints, we have developed a new semi-empirical model for the infrared spectral energy distribution of normal star-forming galaxies between 3 and 1100 $\mu$m.  A sequence of galaxy spectra are constructed from a collection of ``local'' dust emission curves by assuming dust mass is spread over a wide range of interstellar radiation fields according to a power law distribution.  Each ``local'' emission curve corresponds to a different interstellar radiation field heating intensity and is a combination of emission curves for large and very small grains and aromatic feature carriers.   The model reproduces well the empirical spectra and infrared color trends for our sample.  These model spectra allow us to determine the infrared energy budget for normal galaxies, and in particular to translate far-infrared fluxes into total (3 to 1100 \m) infrared fluxes.  The 20 to 42 \m\ range appears to show the most significant growth in relative terms as the activity level increases, suggesting that the 20--42 \m\ continuum may be the best dust emission tracer of current star formation in galaxies.  We have also investigated the redshift dependence of \SIRTF\ infrared color-color diagrams and the far-infrared to radio correlation for normal galaxies.  In our next paper we plan to address the concerns outlined in the discussion section and to further constrain the model outside the 3-100 \m\ range by using 2MASS $JHK$ and submillimeter data.

\acknowledgements 
We have benefited from helpful discussions with F. Boulanger, J. Brauher, B. Draine, S. Malhotra, G. Neugebauer, and B.T. Soifer, and from useful suggestions from the editor, Steven Willner, and the referee.  This work was supported by \ISO\ data analysis funding from the U.S. National Aeronautics and Space Administration, and carried out at the Infrared Processing and Analysis Center and the Jet Propulsion Laboratory of the California Institute of Technology.  \ISO\ is an ESA project with instruments funded by ESA member states (especially the PI countries: France, Germany, the Netherlands, and the United Kingdom), and with the participation of ISAS and NASA.

\scriptsize
\begin{deluxetable}{ccccccccc}
\tablenum{1}
\def\la{$\langle$}
\def\ra{$\rangle$}
\tablewidth{0pt}
\tablecaption{Average Galaxy Infrared Broadband Flux Ratios}
\tablehead{
\colhead{$\Delta$ 60/100} & \colhead{N} & \colhead{\la 6.75/12\ra} & \colhead{\la 6.75/15\ra} & \colhead{\la 6.75/25\ra} & \colhead{\la 6.75/60\ra} & \colhead{\la6.75/100\ra} & \colhead{\la 12/15\ra} & \colhead{\la 12/25\ra}
}
\startdata
$-0.55 \rightarrow -$0.45 & ~8 & $-$0.28(15) & $-$0.13(02) & $-$0.39(07) & $-$1.31(08) & $-$1.79(08) & $+$0.15(15) & $-$0.09(13) \nl
$-0.45 \rightarrow -$0.35 & 11 & $-$0.30(08) & $-$0.12(03) & $-$0.41(14) & $-$1.43(11) & $-$1.83(11) & $+$0.17(10) & $-$0.13(12) \nl
$-0.35 \rightarrow -$0.25 & 16 & $-$0.28(20) & $-$0.10(11) & $-$0.55(21) & $-$1.49(21) & $-$1.80(20) & $+$0.18(19) & $-$0.31(19) \nl
$-0.25 \rightarrow -$0.15 & 19 & $-$0.29(13) & $-$0.21(12) & $-$0.77(24) & $-$1.77(31) & $-$1.98(29) & $+$0.08(13) & $-$0.48(23) \nl
$-0.15 \rightarrow -$0.05 & ~9 & $-$0.43(20) & $-$0.44(20) & $-$1.15(36) & $-$2.06(40) & $-$2.17(37) & $-$0.01(14) & $-$0.71(21) \nl
$-0.05 \rightarrow +$0.05 & ~4 & $-$0.64(26) & $-$0.62(26) & $-$1.50(35) & $-$2.32(52) & $-$2.34(52) & $+$0.03(12) & $-$0.85(19) \nl
$+0.05 \rightarrow +$0.15 & ~2 & $-$0.60(03) & $-$0.73(12) & $-$1.46(21) & $-$2.11(22) & $-$1.99(19) & $-$0.12(10) & $-$0.86(18) \nl
\hline\\
$\Delta$ 60/100 & \la 12/60\ra & \la 12/100\ra & \la 15/25\ra & \la 15/60\ra & \la 15/100\ra & \la 25/60\ra & \la 25/100\ra & \la 60/100\ra \nl \hline
$-0.55 \rightarrow -$0.45 & $-$1.03(10) & $-$1.52(09) & $-$0.26(08) & $-$1.18(08) & $-$1.66(08) & $-$0.95(15) & $-$1.43(15) & $-$0.48(03) \nl
$-0.45 \rightarrow -$0.35 & $-$1.12(10) & $-$1.52(09) & $-$0.29(16) & $-$1.30(12) & $-$1.70(13) & $-$0.99(17) & $-$1.39(16) & $-$0.40(03) \nl
$-0.35 \rightarrow -$0.25 & $-$1.24(24) & $-$1.55(24) & $-$0.45(17) & $-$1.39(17) & $-$1.70(17) & $-$0.93(11) & $-$1.24(11) & $-$0.31(03) \nl
$-0.25 \rightarrow -$0.15 & $-$1.43(37) & $-$1.64(36) & $-$0.56(17) & $-$1.57(26) & $-$1.78(24) & $-$0.95(23) & $-$1.16(23) & $-$0.21(03) \nl
$-0.15 \rightarrow -$0.05 & $-$1.60(27) & $-$1.71(25) & $-$0.71(17) & $-$1.62(21) & $-$1.73(19) & $-$0.89(09) & $-$1.00(09) & $-$0.11(03) \nl
$-0.05 \rightarrow +$0.05 & $-$1.68(40) & $-$1.70(40) & $-$0.88(30) & $-$1.71(51) & $-$1.72(51) & $-$0.83(21) & $-$0.84(21) & $-$0.02(01) \nl
$+0.05 \rightarrow +$0.15 & $-$1.50(20) & $-$1.39(16) & $-$0.73(09) & $-$1.38(10) & $-$1.27(06) & $-$0.65(01) & $-$0.53(02) & $+$0.11(04) \nl
\hline
\enddata
\label{tab:flux_ratios}
\tablenotetext{}{Table values are logarithmic ratios of flux densities $f_\nu$. The 1$\sigma$ scatter (x100) in each ratio is indicated in parentheses.}
\end{deluxetable}
\normalsize

\scriptsize
\begin{deluxetable}{ccccccccccc}
\tablenum{2}
\def\a{\tablenotemark{a}}
\tablewidth{530pt}
\tablecaption{Infrared Energy Budget}
\tablehead{
\colhead{log \IRAScolor} & \colhead{$\alpha$}        & \colhead{$U\a_{\rm global}$}  & 
\colhead{3-5 $\mu$m}     & \colhead{5-13 $\mu$m}     & \colhead{13-20 $\mu$m}        & 
\colhead{20-42 $\mu$m}   & \colhead{42-122 $\mu$m}   & \colhead{122-1100 $\mu$m}     & 
\colhead{PAH/BG}         & \colhead{VSG/BG}
\nl
\colhead{}   & \colhead{}   &\colhead{} & \colhead{\%} & \colhead{\%} & \colhead{\%} & 
\colhead{\%} & \colhead{\%} & \colhead{\%} & \colhead{} & \colhead{}
}
\startdata
$+$0.11 & 1.06 & 97   & 0.4 & ~3.3 & 4.9 & 33.6 & 52.7 & ~5.0 & 0.06 & 0.30 \nl
$+$0.00 & 1.44 & 18   & 0.8 & ~4.9 & 4.3 & 27.6 & 53.5 & ~8.8 & 0.10 & 0.29 \nl
$-$0.10 & 1.66 & 6.6  & 1.2 & ~6.8 & 4.0 & 22.2 & 52.1 & 13.7 & 0.13 & 0.29 \nl
$-$0.20 & 1.81 & 3.5  & 1.6 & ~8.5 & 3.7 & 17.7 & 49.8 & 18.6 & 0.20 & 0.29 \nl
$-$0.29 & 1.97 & 2.0  & 2.0 & 10.2 & 3.5 & 13.7 & 46.4 & 24.3 & 0.25 & 0.28 \nl
$-$0.40 & 2.19 & 1.2  & 2.3 & 11.9 & 3.3 & ~9.4 & 41.2 & 31.8 & 0.30 & 0.28 \nl
$-$0.51 & 2.50 & 0.75 & 2.6 & 13.1 & 3.3 & ~6.7 & 35.2 & 39.2 & 0.34 & 0.28 \nl
\enddata
\label{tab:IR_budget}
\tablenotetext{a}{The global heating intensity for a model galaxy spectrum can be computed by combining the heating intensities assigned to the local spectra in the fashion outlined by Equation \ref{eq:dMdU}:
$U_{\rm global} = {\Sigma \; U M_{\rm d}(U) \over \Sigma M_{\rm d}(U)}.$}
\end{deluxetable}
\normalsize


\newpage
\begin {thebibliography}{dum}
\bibitem[]{}Bekki, K. \& Shioya, Y. 2000, \apj, 542, 201
\bibitem[]{}Bianchi, S., Davies, J.I. \& Alton, P.B. 2000, \aap, 359, 65
\bibitem[]{}Boselli, A., Lequeux, J., Contursi, A. et al. 1997, \aap, 324, L13
\bibitem[]{}Boselli, A., Lequeux, J., Sauvage, M., Boulade, O., Boulanger, F., Cesarsky, D., Dupraz, C., Madden, S., Viallefond, F. \& Vigroux, L. 1998, \aap, 335, 53
\bibitem[]{}Boulanger, F., Beichman, C., D\'{e}sert, F.X., Helou, G., P\'{e}rault, M. \& Ryter, C. 1988, \apjl, 332, L328
\bibitem[]{}Boulade, O. Sauvage, M., Altieri, B. et al. 1996, \aap, 315, L85
\bibitem[]{}de Bruyn, A.G. \& Wilson, A.S. 1976, \aap, 53, 93
\bibitem[]{}Calzetti, D., Armus, L., Bohlin, R.C., Kinney, A.L., Koornneef, J. \& Storchi-Bergmann, T. 2000, \apj, 533, 682
\bibitem[]{}Cesarsky, D., Lequeux, J., Abergel, A., P\'{e}rault, M., Pelazzi, E., Madden, S. \& Tran, D. 1996, \aap, 315, L309
\bibitem[]{}Condon, J.J., Anderson, M.L. \& Helou, G. 1991, \apj, 376, 95
\bibitem[]{}Condon, J.J. 1992, \araa, 30, 575
\bibitem[]{}Contursi, A., Lequeux, J., Hanus, M. et al. 1998, \aap, 336, 662
\bibitem[]{}Contursi, A., Lequeux, J., Cesarsky, D., Boulanger, F., Rubio, M., Hanus, M., Sauvage, M., Tran, D., Bosma, A., Madden, S. \& Vigroux, L. 2000, \aap, in press, astro-ph/0006185
\bibitem[]{}Csabai, I., Connolly, A.J., Szalay, A.S. \& Budav\'ari, T. 1999, \aj, 119, 69
\bibitem[]{}Cutri, R.M., Huchra, J.P., Low, F.J., Brown, R.L. \& vanden Bout, P.A. 1994, \apjl, 424, L65
\bibitem[]{}Dale, D.A., Helou, G., Silbermann, N.A., Contursi, A., Malhotra, S. \& Rubin, R.H. 1999, \aj, 118, 2055
\bibitem[]{}Dale, D.A., Silbermann, N.A., Helou, G., Valjavec, E., Malhotra, S., Beichman, C.A., Brauher, J., Contursi, A., Dinerstein, H.L., Hollenbach, D.J., Hunter, D.A., Kolhatkar, S., Lo, K.Y., Lord, S.D., Lu, N.Y., Rubin, R.H., Stacey, G.J., Thronson, H.A. Jr., Werner, M.W. \& Corwin, H.G. Jr. 2000, \aj, 120, 583
\bibitem[]{}D\'{e}sert, F.X, Boulanger, F. \& Puget, J.L. 1990, \aap, 237, 215 [DBP90]
\bibitem[]{}Devriendt, J.E.G., Guiderdoni, B. \& Sadat, R. 1999, \aap, 350, 381
\bibitem[]{}Draine, B.T. \& Anderson, N. 1985, \apj, 292, 494
\bibitem[]{}Elbaz, D., Cesarsky, C.J., Fadda, D. et al. 1999, \aap, 351, L37
\bibitem[]{}Elvis, M., Wilkes, B.J., McDowell, J.C., Green, R.F., Bechtold, J., Willner, S.P., Oey, M.S., Polomski, E. \& Cutri, R. 1994, \apjs, 95, 1
\bibitem[]{}Fuentes-Williams, T. \& Stocke, J. 1988, \apj, 96, 1235
\bibitem[]{}Garcia, A.M.P. \& Espinosa, J.M.R. 2000; astro-ph/0003349
\bibitem[]{}Geballe, T.R. 1997, in {\it From Stardust to Planetesimals: Contributed Papers}, ed. M.E. Kress, A.G.G.M. Tielens \& Y.J. Pendleton, NASA Conference publication 3343 (Moffett Field: NASA ARC), p.119
\bibitem[]{}Granato, G.L., Lacey, C.G., Silva, L., Bressan, A., Baugh, C.M., Cole, S. \& Frenk, C.S. 2000, \apj, 542, 710
\bibitem[]{}Guiderdoni, B., Hivon, E., Bouchet, F.R. \& Maffei, B. 1998, \mnras, 295, 877
\bibitem[]{}Helou, G., Soifer, B.T. \& Rowan-Robinson, M. 1985, \apjl, 298, L7
\bibitem[]{}Helou, G. 1986, \apj, 311, L33
\bibitem[]{}Helou, G., Khan, I.R., Malek, L. \& Boehmer, L. 1988, \apjs, 68, 151
\bibitem[]{}Helou, G., Ryter, C. \& Soifer, B.T. 1991, \apj, 376, 505
\bibitem[]{}Helou, G., Malhotra, S., Beichman, C.A., Dinerstein, H., Hollenbach, D.J., Hunter, D.A., Lo, K.Y., Lord, S.D., Lu, N.Y., Rubin, R.H., Stacey, G.J., Thronson, Jr., H.A. \& Werner, M.W. 1996, \aap, 315, L157
\bibitem[]{}Helou, G., Becklin, E., Stencel, R.E. \& Wilkes, B. 1997, {\it ASP Conf Ser} 124, 393
\bibitem[]{}Helou, G., Lu, N.Y., Werner, M.W., Malhotra, S. \& Silbermann, N.A. 2000, \apjl, 532, L21
\bibitem[]{}Hudgins, D.M. \& Allamandola, L.J. 1995, J.Phys.Chem., 99, 3033
\bibitem[]{}Hudgins, D.M. \& Allamandola, L.J. 1997, \apjl, 516, L41
\bibitem[]{}Kleinmann, S.G., Hamilton, D., Keel, W.C., Wynn-Williams, C.G., Eales, S.A., Becklin, E.E. \& Kuntz, K.D. 1988, \apj, 328, 161
\bibitem[]{}Lagache, G., Abergel, A., Boulanger, F., D\'{e}sert, F.X. \& Puget, J.L. 1999, \aap, 344, 322
\bibitem[]{}Lu, N.Y. et al. 2000, in preparation
\bibitem[]{}Malhotra, S., Kaufman, M.J., Hollenbach, D.J. et al. 2000, \apj, submitted
\bibitem[]{}Metcalfe, L, Steel, S.J., Barr, P. et al. 1996, \aap, 315, L105
\bibitem[]{}Poggianti, B.M. \& Wu, H. 2000, \apj, 529, 157
\bibitem[]{}Pozzetti, L., Madau, P., Zamorani, G., Ferguson, H.C. \& Bruzual G.A. 1998, \mnras, 298, 1133)
\bibitem[]{}Roelfsema, P.R., Cox, P., Tielens, A.G.G.M. et al. 1996, \aap, 315, L289
\bibitem[]{}Rowan-Robinson, M., Lawrence, A., Oliver, S., Taylor, A., Broadhurst, T.J., McMahon, R.G., Benn, C.R., Condon, J.J., Lonsdale, C.J., Conrow, T., Saunders, W.S., Clements, D.L., Ellis, R.S. \& Robson, I. 1993, \mnras, 261, 513
\bibitem[]{}Rowan-Robinson, M., Mann, R.G., Oliver, S.J. et al. 1997, \mnras, 289, 490
\bibitem[]{}Rowan-Robinson, M. 2000, \mnras, astro-ph/9912286
\bibitem[]{}Rush, B., Malkan, M., Fink, H.H. \& Voges, W. 1996, \apj, 471, 190
\bibitem[]{}Sanders, D.B. \& Mirabel, I.F. 1996, \araa, 34, 749
\bibitem[]{}Siebenmorgan, R., Kr\"{u}gel, E. \& Chini, R. 1999, \aap, 351, 495
\bibitem[]{}Silva, L., Granato, G.L., Bressan, A. \& Danese, L. 1998, \apj, 509, 103
\bibitem[]{}Steidel, C.C., Giavalisco, M., Pettini, M., Dickinson, M. \& Adelberger, K. 1996, \apjl, 462, L17
\bibitem[]{}Sturm, E., Lutz, D., Tran, D., Feuchtgruber, H., Genzel, R., Kunze, D., Moorwood, A.F.M. \& Thornley, M. 2000, \aap, 358, 481
\bibitem[]{}Telesco, C.M. 1993, \mnras, 263, L37
\bibitem[]{}Tokunaga, A.T. 1997, in {\it From Stardust to Planetesimals: Contributed Papers}, ed. M.E. Kress, A.G.G.M. Tielens \& Y.J. Pendleton, NASA Conference publication 3343 (Moffett Field: NASA ARC)
\bibitem[]{}Tran, D. 1998, Ph.D. Thesis, Universit\'{e} de Paris
\bibitem[]{}Uchida, K., Sellgren, K. \& Werner, M.W. 1998
\bibitem[]{}Verstraete, L., Puget, J.L., Falgarone, E., Drapatz, S., Wright, C.M. \& Timmerman, R. 1996, \aap, 315, L337
\bibitem[]{}Vigroux, L., Mirabel, F., Altieri, B. et al. 1996, \aap, 315, L93
\bibitem[]{}Vigroux, L., Charmandaris, V., Gallais, P., Laurent, O., Madden, S., Mirabel, F., Roussel, H., Sauvage, M. \& Tran, D. 1999, in {\it The Universe as seen by ISO}, eds. P. Cox \& M.F. Kessler (ESA SP-427, France), Vol. 2, p.805
 
\end {thebibliography}
\end{document}